\documentstyle[epsfig,12pt]{article}

\def\Tr{\hbox{Tr}}

\def\Ket#1{|#1\rangle}

\begin{document}


\begin{titlepage}
\begin{centering}
{\Large\bf Two-dimensional super Yang--Mills theory
investigated with improved resolution}\\
\vspace*{0.5cm}

{\bf John R.~Hiller}
\vspace*{0.3cm}

{\sl Department of Physics \\
University of Minnesota Duluth\\
Duluth, MN  55812} \vspace*{0.5cm}

{\bf  Motomichi Harada, Stephen Pinsky, Nathan Salwen}
\vspace*{0.3cm}

{\sl Department of Physics \\
Ohio State University\\
Columbus, OH 43210}\vspace*{0.5cm}

{\bf  Uwe Trittmann}
\vspace*{0.3cm}

{\sl Department of Physics and Astronomy\\
Otterbein College\\
Westerville, OH 43081}

\vspace*{0.3cm}
November 23, 2004
\vspace*{0.3cm}


\begin{abstract}
In earlier work, ${\cal N}=(1,1)$ super Yang--Mills theory in
two dimensions was found to have several
interesting properties, though these properties
could not be investigated in any detail. In this paper we
analyze two of these properties. First, we investigate the
spectrum of the theory. We calculate the masses of
the low-lying states using the supersymmetric discrete light-cone
(SDLCQ) approximation and obtain their continuum
values. The spectrum exhibits an interesting distribution of masses,
which we discuss along with a toy model for this pattern. We also discuss
how the average number of partons grows in the bound states.
Second, we determine the number of fermions and bosons in the
${\cal N}=(1,1)$ and ${\cal N}= (2,2)$ theories in each
symmetry sector as a function of the resolution. Our finding that the
numbers of fermions and bosons in each sector are the same is part of 
the answer to the question of why the SDLCQ approximation exactly
preserves supersymmetry.
\end{abstract}
\end{centering}

\vfill

\end{titlepage}
\newpage

\section{Introduction}
Over the last several years we have solved numerically a number
of supersymmetric theories~\cite{Antonuccio:1999zu,Haney:2000tk,hpt2001}.
The simplest of these is ${\cal N}=(1,1)$ supersymmetric
Yang--Mills (SYM) theory in
$1+1$ dimensions with a large number of colors, $N_c$. We have
investigated this theory
previously~\cite{Antonuccio:1998kz,Hiller:2004vy}
but have not studied its properties in detail at high resolution. This is
the purpose of the present paper.

We use a supersymmetric version of Discrete Light-Cone
Quantization (SDLCQ)~\cite{sakai,Lunin:1999ib} to solve this theory. 
The SDLCQ approach is described in a number of other places,
and we will not present a detailed discussion of
the method here; for a review, see~\cite{Lunin:1999ib}.
For those readers familiar
with DLCQ~\cite{pb85,bpp98}, it suffices to say that both methods
use light-cone coordinates, with $x^\pm\equiv(x^0\pm x^1)/\sqrt{2}$, and
have discrete momenta and cutoffs in momentum space.
In two dimensions the discretization is specified by a
single integer $K=(L/\pi) P^+$, the harmonic resolution~\cite{pb85}, such
that the longitudinal
momentum fractions are integer multiples of $1/K$. SDLCQ is
formulated in such a way that it preserves supersymmetry
in each step of the calculation.  

Exact supersymmetry brings a number of
very important numerical advantages to the method. In particular,
theories with enough supersymmetry are finite. We have also seen
greatly improved numerical convergence in this approach.  Currently,
SDLCQ seems to be the only method available for numerically solving 
strongly coupled SYM theories.  Conventional lattice methods have
difficulty with supersymmetric theories because of the asymmetric
way that fermions and bosons are treated, and
progress~\cite{lattice} in supersymmetric lattice gauge theory has
been relatively slow.

In this paper we will focus on two aspects of the solution of
${\cal N}=(1,1)$ SYM theory that have been noted previously but never
investigated. The first is the mass spectrum of the theory. In
earlier work~\cite{Antonuccio:1998kz} it was found that at finite
numerical resolution this theory
has a mass gap. However, as the resolution increased, the
gap closed linearly as a function of the inverse resolution. The
lighter states which appeared in the gap at higher resolution have
more partons. This is in fact the reason why these states are not seen at
lower resolution. Obviously, at resolution $K$ in the SDLCQ approximation
one can see states with at most $K$ partons.
States with a dominant contribution of more than $K$ partons will be
missed.  Thus there is a curve
$M^2= a +b/K$ that predicts the mass at which a bound state first appears
in a mass versus resolution plot.
Of course, the state then consistently appears at each higher $K$.
To find the
continuum mass of a particular bound state, one fits its mass as a
function of the resolution and extrapolates to infinite $K$. We
will perform this calculation and find the continuum masses of a
number of states. These fits turn out to be quite flat and vary only 
weakly with the resolution. One of the characteristics of the SDLCQ 
formulation of ${\cal N}=(1,1)$ SYM theory is that the mass
at initial appearance of the state is relatively close to the continuum
mass. The detailed study of the theory we present here
will show that there are in fact many
sequences of these curves from which states emerge. One might speculate
that the entire spectrum can be understood in terms of these curves of
first appearance.

The second aspect is development of
a formula to determine the number of states in
each distinct sector of a given theory without explicitly counting them.
By comparing these numbers of states in different sectors we can
immediately determine a lower bound on 
the number of massless states.  The general result is that the total 
number of states increases as $(n+1)^K$, where $n$ is the
number of types of particles and $K$ is the resolution.  For
${\cal N}=(1,1)$ SYM theory, the number of states increases as $3^K$.
We also show that the numbers of fermion and boson Fock states are equal
for each parton number. The fact that in each sector
the number of fermionic states equals the number of
bosonic states, for all theories considered, is part of the answer to the
fundamental question of why SDLCQ exactly preserves supersymmetry.

A third aspect remains to be fully analyzed; this is
is the structure of the large set of massless states.
These states can be arranged so as to be nearly pure
states of a fixed number of partons. An investigation of
the Fock-state content is underway to determine whether
the small deviation from being a pure parton-number state is a numerical
artifact or a fundamental property of the theory.

The material in the present paper is organized as follows.
In Sec.~\ref{sec:Super} we briefly review SYM theory
in 1+1 dimensions.  The discussion in
Sec.~\ref{sec:spectrum} describes the spectrum and 
our numerical calculation of the
properties of the bound states.  In
Sec.~\ref{sec:prime}, we give the calculation of the number of states
when the harmonic resolution $K$ is prime,
while Sec.~\ref{sec:generalk} outlines the
calculation for general $K$. We then go on to
calculate the number of states in each of the
symmetry sectors in
Sec.~\ref{sec:symmetry}. In Sec.~\ref{sec:discussion}, we discuss our
results.

\section{Super Yang--Mills theory}
\label{sec:Super}
We will start by providing a brief review of ${\cal N}=(1,1)$
supersymmetric Yang--Mills theory in 1+1 dimensions. The Lagrangian of
this theory is
\begin{equation}
{\cal L}
 ={\rm Tr}\left(-\frac{1}{4}F_{\mu\nu}F^{\mu\nu}
      +i\bar{\Psi}\gamma_{\mu}D^{\mu}\Psi\right).\label{Lagrangian}
\end{equation}
The two components of the spinor $\Psi=2^{-1/4}({\psi \atop
\chi})$ are in the adjoint representation, and we will work in the
large-$N_c$ limit.  The field strength and the covariant
derivative are
$F_{\mu\nu}=\partial_{\mu}A_{\nu}-\partial_{\nu}A_{\mu}
+ig[A_{\mu},A_{\nu}]$ and $D_{\mu}=\partial_{\mu}+ig[A_{\mu}]$.
The most straightforward way to formulate the theory in 1+1
dimensions is to start with the theory in 2+1 dimensions and then
simply dimensionally reduce to 1+1 dimensions by setting
$\phi=A_2$ and $\partial_2 \rightarrow 0$ for all fields.  In the
light-cone gauge, $A^+=0$, we find
\begin{equation}
Q^-=2^{3/4}g\int dx^- \left(i[\phi,\partial_-\phi]
         +2\psi\psi\right)\frac{1}{\partial_-}\psi .
\end{equation}
The mode expansions in two dimensions are
\begin{eqnarray}
\phi_{ij}(0,x^-) &=& \frac{1}{\sqrt{2\pi}} \int_0^\infty
         \frac{dk^+}{\sqrt{2k^+}}\left[
         a_{ij}(k^+)e^{-{\rm i}k^+x^-}+
         a^\dagger_{ji}(k^+)e^{{\rm i}k^+x^-}\right] ,
\nonumber\\
\psi_{ij}(0,x^-) &=&\frac{1}{2\sqrt{\pi}}\int_0^\infty
         dk^+\left[b_{ij}(k^+)e^{-{\rm i}k^+x^-}+
         b^\dagger_{ji}(k^+)e^{{\rm i}k^+x^-}\right] .
\end{eqnarray}

To obtain the spectrum, we solve the mass eigenvalue problem
\begin{equation}
2P^+P^-|\varphi\rangle=M^2|\varphi\rangle .
\label{EVP}
\end{equation}
We convert this to a matrix eigenvalue problem by introducing a
discrete basis where $P^+$ is diagonal.
In SDLCQ this is done by first discretizing the supercharge
$Q^-$ and then constructing
$P^-$ from the square of the supercharge: $P^- =
(Q^-)^2/\sqrt{2}$.  To discretize the theory we impose periodic
boundary conditions on the boson and fermion fields alike and
obtain an expansion of the fields with discrete momentum modes. We
define the discrete longitudinal momenta $k^+$ as fractions
$nP^+/K$ of the total longitudinal momentum $P^+$, where the integer $K$
determines the resolution of the discretization. Because
light-cone longitudinal momenta are always positive, $K$ and each
$n$ are positive integers; the number of partons is then
bounded by $K$.   The continuum limit is recovered by taking the
limit $K \rightarrow \infty$.

In constructing the discrete approximation we drop the
longitudinal zero-momentum modes.  For some discussion of dynamical
and constrained zero modes, see the review~\cite{bpp98} and
previous work~\cite{alpt98}. Inclusion of these modes would be
ideal, but the techniques required to include them in a numerical
calculation have proved to be difficult to develop, particularly
because of nonlinearities.   For DLCQ calculations that can be
compared with exact solutions, the exclusion of zero modes does
not affect the massive spectrum~\cite{bpp98}. In scalar theories
it has been known for some time that constrained zero modes can
give rise to dynamical symmetry breaking~\cite{bpp98}, and work
continues on the role of zero modes and near zero modes in these
theories~\cite{Rozowsky:2000gy,Salmons:2002xg,Heinzl:2003jy,%
Harindranath:ud,Chakrabarti:2003ha}.

The ${\cal N}=(1,1)$ SYM theory has two discrete symmetries besides
supersymmetry that
we use to block diagonalize the Hamiltonian matrix. $S$-symmetry,
which is associated with the orientation
of the large-$N_c$ string of partons in a state~\cite{kut93},
gives a sign when the color indices are permuted
\begin{equation}\label{Z2-S}
S : a_{ij}(k^+)\rightarrow -a_{ji}(k^+) , \qquad
      b_{ij}(k^+)\rightarrow -b_{ji}(k^+) .
\end{equation}
$P$-symmetry is what remains of parity in the $x^2$ direction
after dimensional reduction
\begin{equation}\label{Z2-P}
P : a_{ij}(k^+)\rightarrow -a_{ij}(k^+), \qquad
      b_{ij}(k^+)\rightarrow b_{ij}(k^+).
\end{equation}
All of our states can be labeled by the $P$ and $S$ sector in
which they appear.


\section{The spectrum and properties of states}
\label{sec:spectrum}

\subsection{The mass spectrum}

The massive spectrum of this theory has a nontrivial pattern that can be
analyzed and understood. In this section we will present the results of
our numerical calculation of this spectrum, describe the pattern, and 
present an analytic model of it. There are two distinct sectors of
the theory: one comprises the states even under the $S$-symmetry and
the other sector the ones odd under this operation. Additionally,
the masses in each sector are degenerate because of parity and
supersymmetry. We will analyze the two sectors separately.

First we need to describe how a bound state appears in the spectral data.
If we plot the mass (squared)\footnote{Note that we consistently write
$M^2$ in units of $g^2 N_c/\pi$.}
of the states as a function of the inverse harmonic resolution $1/K$,
a bound state is represented by a sequence of points starting at some
minimum resolution $K_0$, appearing at every higher $K$ with slightly 
different mass to converge to a continuum mass as $K\rightarrow \infty$.
That the state is not seen until some minimum resolution $K_0$ is reached
reflects the fact that the dominant number
of partons is $K_0$. Thus at resolutions less than $K_0$ we simply cannot
have such a state.
At resolutions above $K_0$, we see better and better representations of
the state.  Since a state has similar properties at resolutions $K$ and 
$K+1$, we can follow the sequence of masses in this string of states 
with increasing resolution and extrapolate to infinite resolution, 
yielding the continuum mass of the state. 

\begin{figure}[t]
\begin{center}
\begin{tabular}{cc}
\psfig{figure=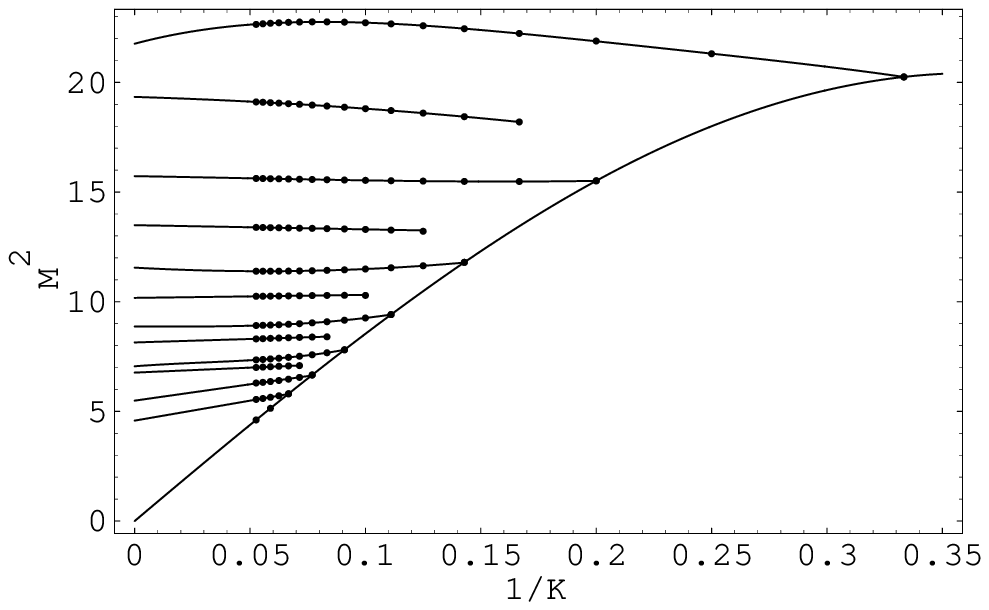,width=7cm}&
\psfig{figure=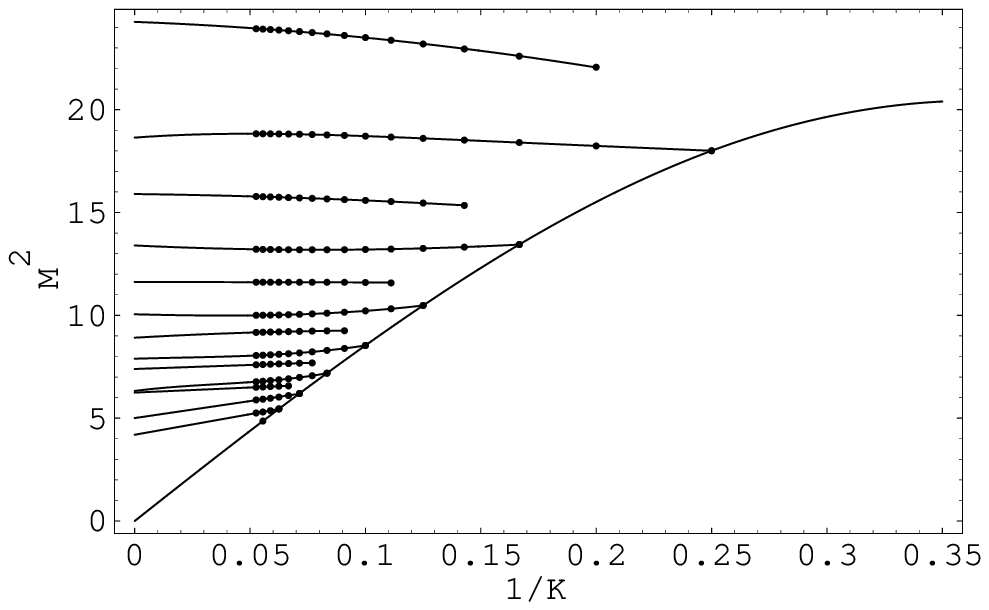,width=7cm}
\\
(a)&(b)
\end{tabular}
\end{center}
\caption{The mass spectrum for (a) $S=+1$ and (b) $S=-1$, as a function
of the inverse resolution $1/K$. Shown are polynomial fits to the data
and a fit to the lowest state.}
\label{SpectrumFits}
\end{figure}

The structure of the spectrum is best understood by starting with the
low-lying masses at low resolution. If we see a sequence of states
starting at
some harmonic resolution $K_0$, the next new lowest state in the scheme
will appear at resolution $K_0+2$ and will have a smaller
mass. For example, in the $S$-even sector,
we see one state at resolution $K=3$. We do not
see another new state until resolution $K=5$. Comparing the two $S$
sectors, in Fig.~\ref{SpectrumFits}, we  see that
$K_0$ is even in the $S$-odd sector and vice versa. We thus have a unique
naming scheme for the states by naming them after the resolution
$K_0$ of their first appearance on the plot.
A fit to the masses as a function of the harmonic
resolution with (at most) cubic polynomials gives an unambiguous and
clean picture. The continuum values ($K\rightarrow\infty$) are easily
extracted.  We obtain the values on the left side of
Tables~\ref{MassTable0}
and \ref{MassTable1}. These states are characterized by the value $K_0$.
For example, the mass of a particular bound
state is $M^2_{K_0}(K)$, its continuum value is $M^2_{K_0}(\infty)$, and
the mass of the bound state when it first appears is $M^2_{K_0}(K_0)$.
The states that appear at high values of $K_0$ have more partons and are
lighter.  If we think of the trace of partons as a string, then as a 
general rule the longer strings are lighter.
\begin{table}
\centerline{
\begin{tabular}{cccc}\hline \hline
 $K_0$ & $M^2_{K_0}(\infty)$ & $K_1$ & $M^2_{K_1}(\infty)$ \\ \hline
   3  & 21.7644 &  &  \\
  5 & 15.7268 & 6& 19.3398 \\
  7& 11.5539 & 8 & 13.5907 \\
  9&  8.87045 & 10& 10.1742\\
  11& 7.0556 & 12& 8.14241 \\
  13 &5.48951 & 14 & 6.76741\\
  15 &4.57762 & & \\
   \hline \hline
\end{tabular}}
\caption{Mass-squared eigenvalues in the $S$-even sector. Displayed
are the continuum masses $M^2_{K_n}(\infty)$ of the states appearing
at $K_0$ (left) and $K_1$ (right).}
\label{MassTable0}
\end{table}
\begin{table}
\centerline{
\begin{tabular}{cccc}\hline \hline
 $K_0$ & $M^2_{K_0}(\infty)$ & $K_1$ & $M^2_{K_1}(\infty)$ \\ \hline
  4  & 18.6402 & 5 & 24.2683\\
  6 & 13.3970 & 7& 15.8981  \\
  8& 10.0517 & 9& 11.6391\\
  10& 7.89029 & 11&  8.91274\\
  12& 6.32305 & 13&7.39131 \\
  14 & 5.00495 & 15& 6.24411\\
  16 & 4.19408 & & \\
   \hline \hline
\end{tabular}}
\caption{Mass-squared eigenvalues in the $S$-odd sector. Displayed
are the continuum masses $M^2_{K_n}(\infty)$ of the states appearing
at $K_0$ (left) and $K_1$ (right).}
\label{MassTable1}
\end{table}

\begin{figure}[t]
\begin{center}
\begin{tabular}{cc}
\psfig{figure=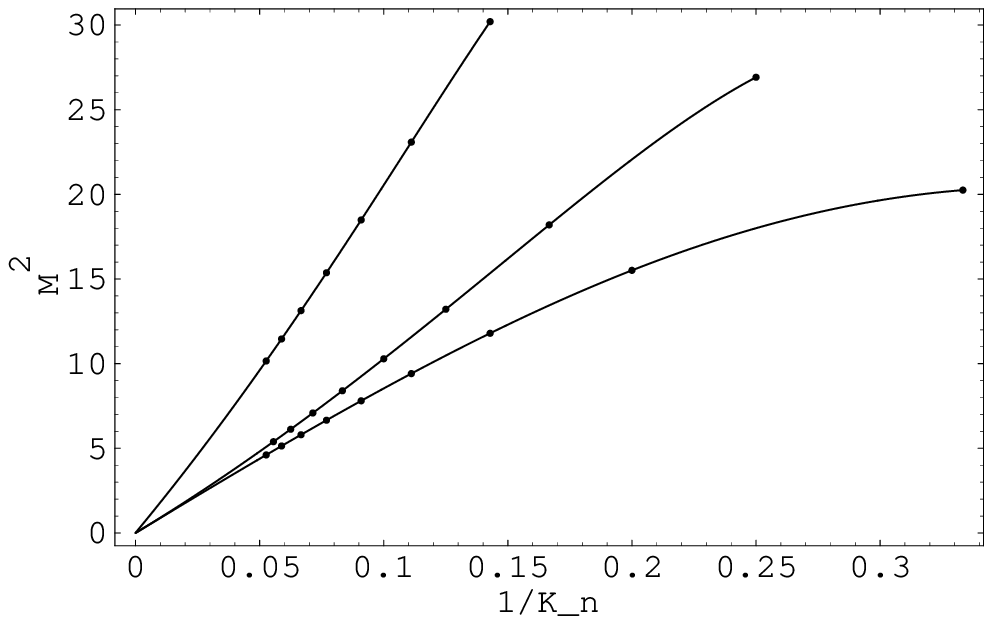,width=7cm}&
\psfig{figure=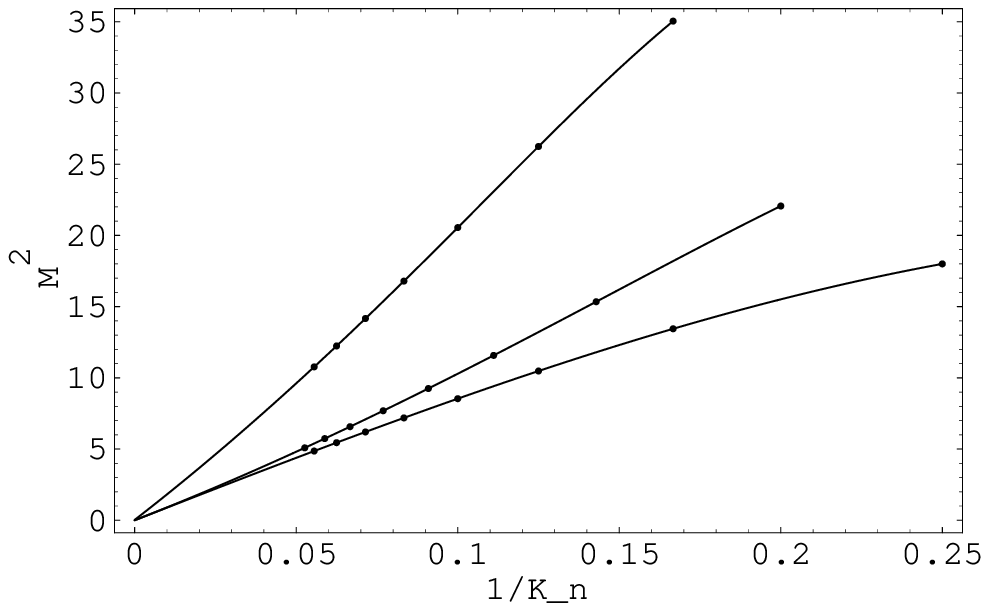,width=7cm}
\\
(a)&(b)
\end{tabular}
\end{center}
\caption{The sequences $M^2_{K_n}(K_n)$
of low-mass states for (a) $S=+1$ and (b) $S=-1$, as functions
of the inverse resolution $1/K_n$. Shown are the masses of
the sequences with $K_0=3,5,7,\ldots$, $K_1=4,6,8,\ldots$, 
$K_2=7,9,11,\ldots$ in the 
$S$-even sector, and at those that begin at $K_0=4$, $K_1=5$, $K_2=6$ in the 
$S$-odd sector. The curves are polynomial fits to the data.}
\label{ZeroFits}
\end{figure}

Consider $M^2_{K_0}(K_0)$, i.e. the mass at which a sequence of 
bound states are first seen, as a function of $K_0$. At even $S$ we have
$K_0=3,5,7,\ldots$ 
and in the $S$-odd sector $K_0$ takes on even integer values starting at four.
The points that make up these curves are shown in Fig.~\ref{ZeroFits}
along with the following polynomial fits:
\begin{equation}
M^2_{K_0,S+}(K_0)= -0.000316166 + 88.8448 \frac{1}{K_0} -
9.84279\frac{1}{K_0^2}\pm\ldots,
\end{equation}
\begin{equation}
M^2_{K_0,S-}(K_0)=0.000125688 + 88.8097 \frac{1}{K_0} -8.68279 \frac{1}{K_0^2}
\pm
\ldots,
\end{equation}
\begin{equation}
M^2_{K_0, {\rm all}}(K_0)=
-0.000353165 + 88.8513 \frac{1}{K_0} -10.1359 \frac{1}{K_0^2} \pm
\ldots .
\end{equation}
The curves pass through zero, and
interestingly the $S$-even and $S$-odd curves nicely fit together into a
single curve. The fact that all the curves converge to zero at 
$K_0\rightarrow\infty$ indicates that the mass gap in this theory closes.

Slightly more massive states appear at resolutions $K_1=K_0+1$.  These
sequences and their continuum limits lie between those that begin at 
$K_0$ and $K_0-2$.  We therefore have one such new state for
every state in the $K_0$ series of states. Each of the states
$M^2_{K_1}(K)$
can also be followed as the resolution goes to infinity and the continuum
masses, $M^2_{K_1}(\infty)$, are displayed on the right of
Tables~\ref{MassTable0} and \ref{MassTable1}.

In addition to these two sets of bound states, there are bound states
that appear at masses larger than $M^2_3(3)$ for $S$ even and $M^2_4(4)$ 
for $S$ odd.  They are not shown in Fig.~\ref{SpectrumFits}. We call their
initial resolution $K_2$, and remark that there appears to be a whole
string of states with starting points $K_n$. In Fig.~\ref{ZeroFits}
we have plotted $M^2_{K_n}(K_n)$ versus $1/K_n$ for $n=0,1,2$.
We emphasize that the mass gap closes in all three cases.

In an unrelated matter, we found that the eigenvalues of states of
similar masses show no sign of interaction at all. In an approximation scheme
such as SDLCQ we would expect eigenvalue repulsion due to discretization
artifacts.  None of these $1/K^r$ effects are discernible here.

\subsection{A toy model}

It is instructive to
consider an analytical toy model for a massive spectrum that models the
properties we discussed in the previous section. This model is
simpler than the data we have presented but does provide an
analytic form against which to compare our numerical results.
The properties that characterize the toy model are easy to describe.
First, there is a function
$M^2_{K_0}(K)$ that models the mass at which a bound state first
appears. It is
$M^2_{K_0}=84/K_0$, and, as in the data, the toy spectrum has a mass
gap that closes.  We consider in our
model only the masses less than the lowest mass at
$K=6$, that is $M^2 = 14$. For resolutions less than $K=6$, there are no
states in this model, just as for the $S-$ sector of SYM theory where,
in Table~\ref{MassTable1}, there are no states with $K$ less than four.

The model properties for the data in our model for the massive bound
state in this region are as follows:
\begin{itemize}
\item At resolution $K=6$ the only bound state has $M^2=14$.
\item When we increase the resolution $K$, the spectrum has all
the states that were present at the lower resolution, and their
masses are independent of $K$.
\item When $K$ is increased, there are additional states
that appear between the states that existed at the lower
resolution.
\end{itemize}

\begin{figure}
\centerline{\psfig{figure=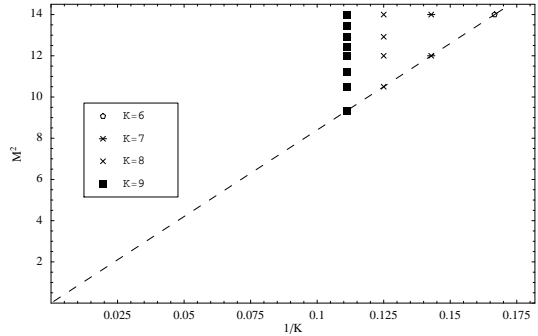,width=7.5cm}}
\caption{The lowest mass squared as a
function of $1/K$ in the toy model. } \label{minmass}
\end{figure}

The spectrum for this toy model, up to resolution K=9, is shown in
Fig.~\ref{minmass}.
At $K=7$ there are states at $M^2=14=84/6$
and 84/7; at $K=8$ there are states at $M^2=14$, 84/6.5, 84/7, and
84/8. It is straightforward to calculate
the number of states with a mass squared less than any arbitrary $M^2$.
We find
\begin{equation}
N(M^2, K) =2^{(K-\frac{84}{M^2})} \Theta(14-M^2) \Theta(K-6).
\end{equation}
Thus at
$K=6$ there is one massive state in the region $M^2 \leq 14$, while at
$K=8$ there are 4
states with $M^2 \leq 14$. The total number of states at any $K$ is
\begin{equation}
N(K)=2^{(K-6)} \Theta(K-6) .
\end{equation}

The density of states can be calculated from $N(M^2,K)$ and is
\begin{equation}
\rho(M^2,K)=\frac{dN(M^2,K)}{dM^2}= N(K) \tilde{\rho}(M^2),
\end{equation}
where the reduced density of states~\cite{Lunin:2000im} $\tilde{\rho}$ is
just
\begin{equation}
\tilde{\rho}(M^2)
    =2^{(6-\frac{84}{M^2})} \frac{84}{M^4} \ln(2) \Theta(14-M^2).
\end{equation}
Clearly this is a peaked function that vanishes at $M^2=0$ and $\infty$.
The location of the peak is set by the maximum mass allowed in the model,
$M^2=14$. Of course one could rewrite the model for a very large maximum
mass in order to model a
larger range of masses. One finds that for masses much less than the mass
at the peak of the density, it grows exponentially with $M$, 
suggesting that the model might have a 
Hagedorn temperature. The thermodynamics of a model displaying a
Hagedorn temperature are discussed in~\cite{Hiller:2004vy}.

\subsection{Properties of the states}
\label{sec:properties}

Since we solve for the wave functions in addition to the masses of the
bound states, we can investigate the properties of the bound states 
discussed above. We already mentioned that the average number of partons
in a bound state is higher for lower mass states. There is a four-fold 
degeneracy for all of the
massive states, and the properties of these degenerate states can be very
different.  Here, we will look at properties such as the average number
of partons and average number of fermions and their dependence on the
symmetry sector to which the bound state belongs.
The first useful observation is that the eight sectors of the theory
contain different numbers of states. It is also interesting to note that
different symmetry
sectors have different numbers of massless states, and that the
average parton and fermion numbers depend on the symmetry sector,
as can be seen in Table~\ref{AvgnTable}.
\begin{table}
\centerline{
\begin{tabular}{ccccccccc}\hline \hline
$S$ & \multicolumn{4}{c}{$S+$ } &
      \multicolumn{4}{c}{$S-$ }   \\
$B$ & \multicolumn{2}{c}{$B+$} & \multicolumn{2}{c}{$B-$}
 & \multicolumn{2}{c}{$B+$} & \multicolumn{2}{c}{$B-$}\\
$P$& $P+$ & $P-$  & $P+$ & $P-$  & $P+$ & $P-$  & $P+$ & $P-$  \\
Massless states? & yes & no & no & yes & no & yes & yes & no \\
\hline
$K_0$: $\langle n \rangle$ & 7.94 & 8.96 & 8.97 & 7.94  & 7.93 & 6.90 &
6.90
& 7.93\\
$K_0$: $\langle n_f \rangle$ & 1.98 & 2.00 & 2.98 & 1.00 & 2.00 & 1.76 &
1.00& 2.79\\
$K_1$: $\langle n \rangle$ & 6.00 & 6.99 & 6.99 & 6.00  & 7.99 & 6.99 &
6.99
& 7.99\\
$K_1$: $\langle n_f \rangle$ & 0.33 & 2.00 & 1.28 & 1.02 & 2.00 & 0.21 &
1.01& 1.19\\
  \hline \hline
\end{tabular}}
\caption{Average parton number $\langle n \rangle$ and fermion number
$\langle n_f \rangle$ in different sectors  at resolution $K=9$, for the
states in the sequences that begin at resolutions $K_0$ and $K_1$.
Here $B+$ and $B-$ indicate the bosonic and fermionic sectors.}
\label{AvgnTable}
\end{table}

Looking at the zero-mass states, we find that they have either
fermion number one or zero. Furthermore, we have states of very different
length, ranging from the smallest to the largest possible number of
partons in each sector.

\begin{figure}[t]
\begin{center}
\begin{tabular}{cc}
\psfig{figure=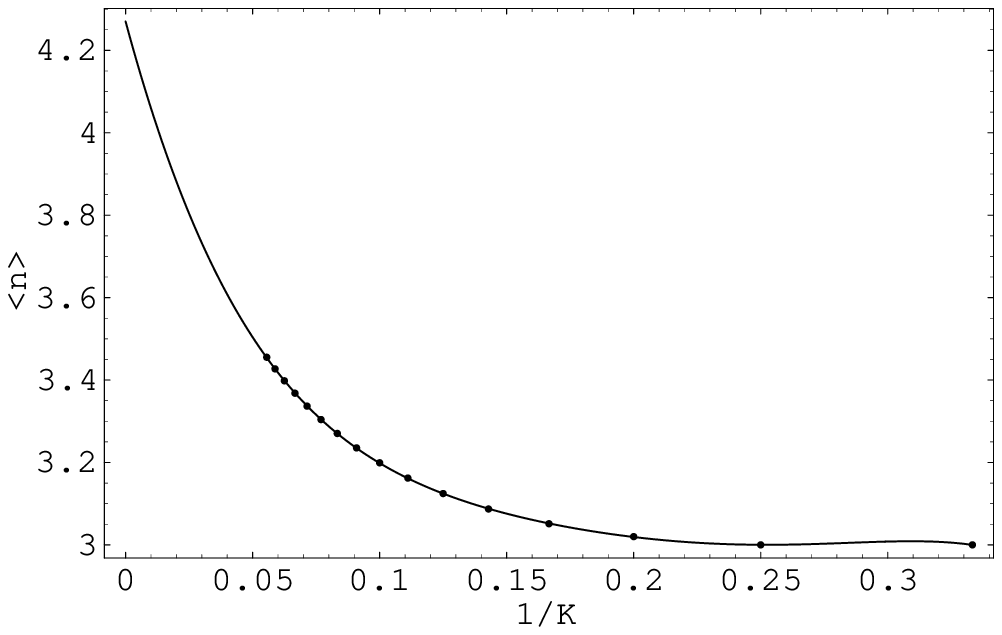,width=7cm} &
  \psfig{figure=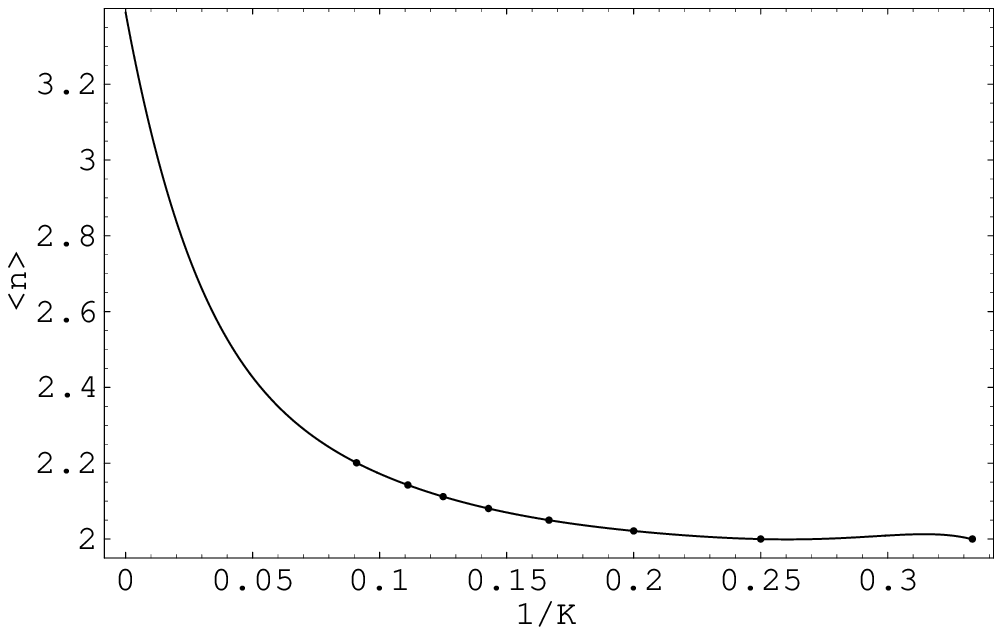,width=7cm}
  \\
(a)&(b)
\end{tabular}
\end{center}
\caption{Average parton number of the first massive state appearing at
$K_0=3$ as a function of $K$ in (a) the bosonic $P-$, $S+$ sector 
(no zero-mass states) and
(b) the bosonic $P+$, $S-$ sector (with zero-mass states). The lines are
polynomial fits to the data.\label{avgn}}
\end{figure}

It is interesting to consider the parton number of a bound state as a
function of the harmonic resolution. As an example, we plot in
Fig.~\ref{avgn}
the average parton number of the lowest massive state that first
appears at $K=3$ .
We see that the parton number grows with $K$ in
the sector with zero-mass states as well as in the sector without them.

\begin{figure}[t]
\begin{center}
\begin{tabular}{cc}
\psfig{figure=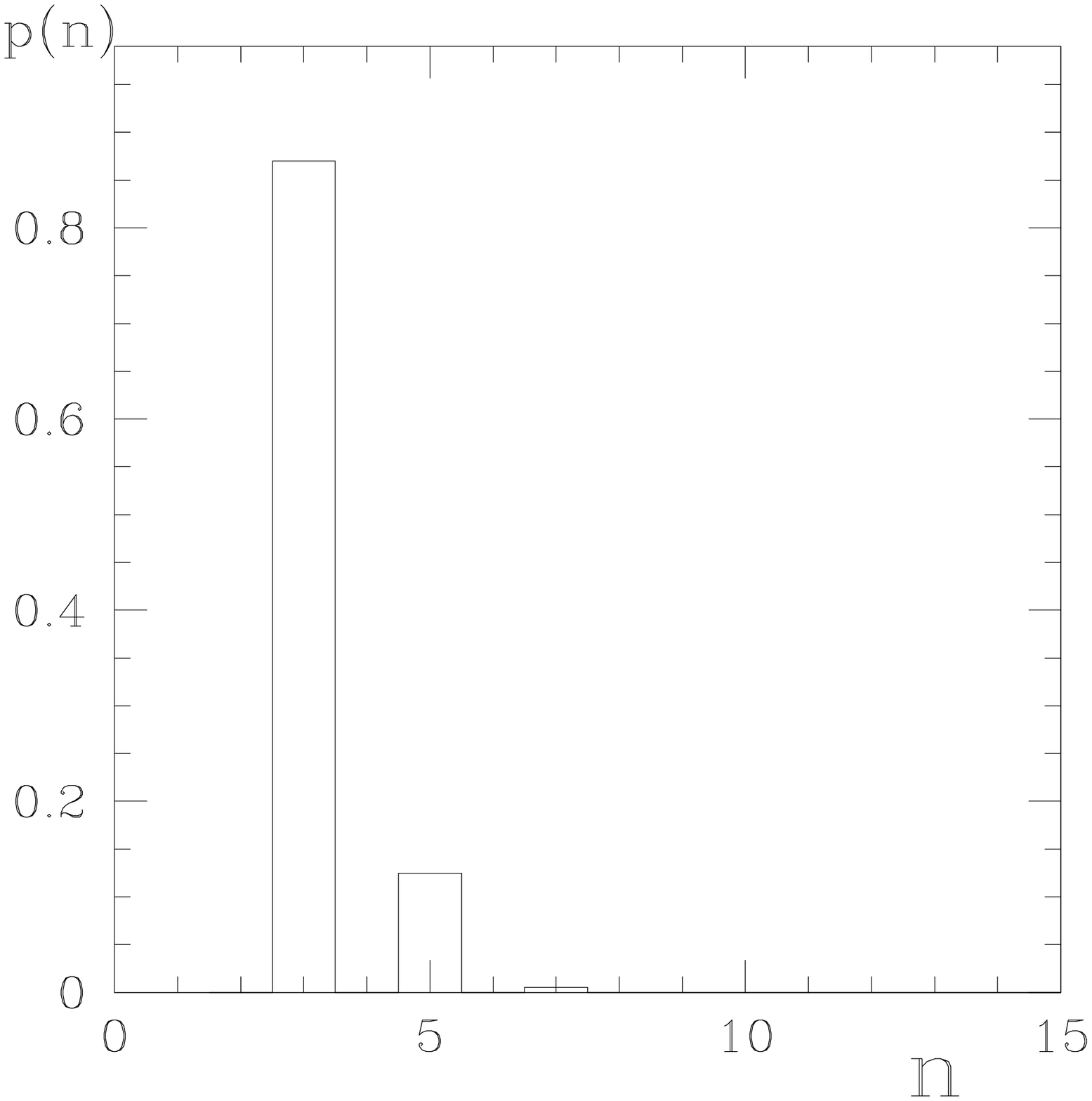,width=5.7cm} &
            \psfig{figure=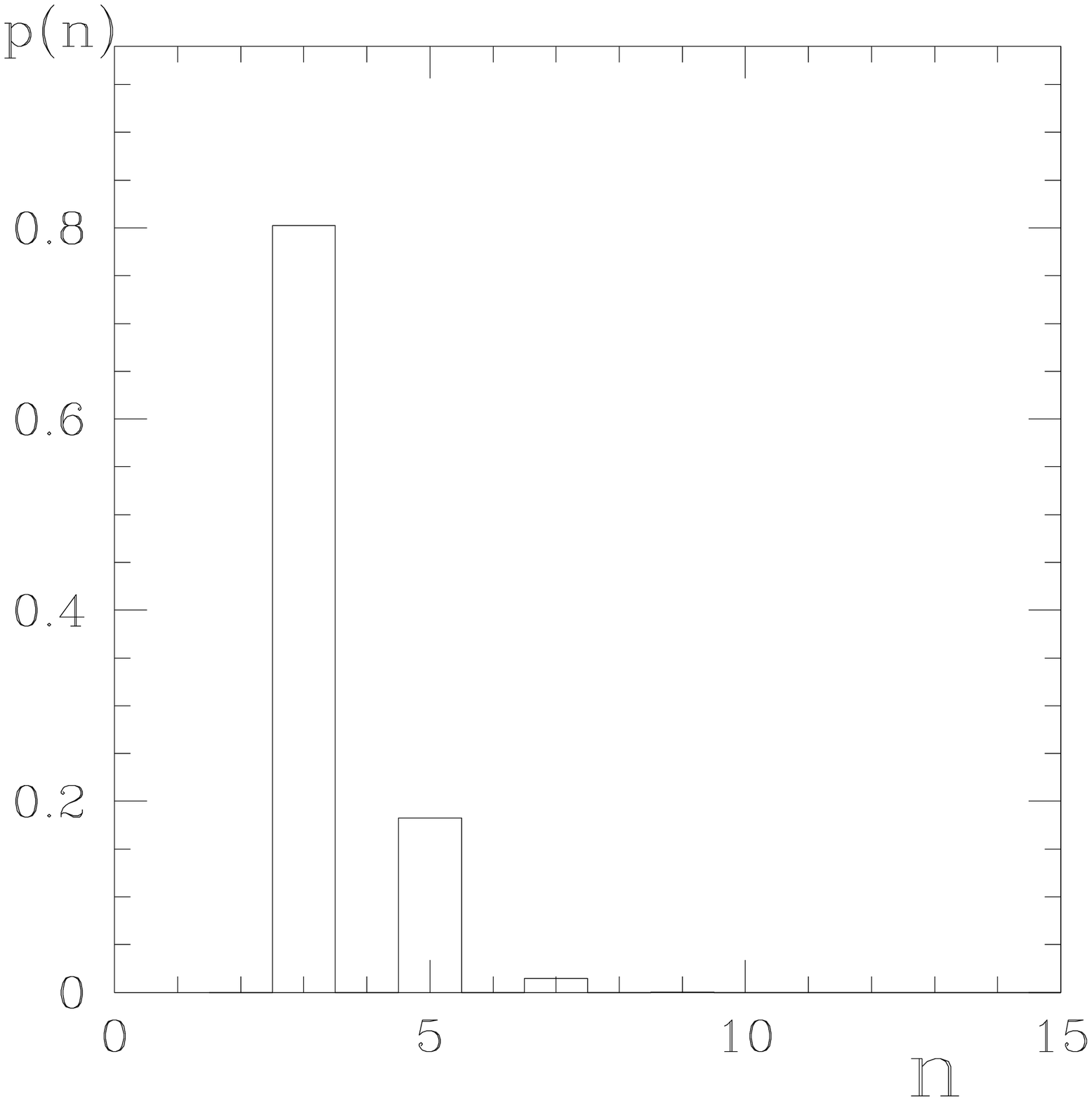,width=5.7cm}
  \\
(a)&(b)
\end{tabular}
\end{center}
\caption{Parton number distribution for the first state
appearing in the spectrum ($M_3^2(\infty)=21.7644$ in the bosonic $P-$,
$S+$ sector).
Plotted are the probabilities $p(n)$ for the state to have $n$ partons vs $n$,
for resolutions (a) $K=12$ and (b) $K=17$.
\label{ppd}}
\end{figure}

\begin{figure}[t]
\centerline{\psfig{figure=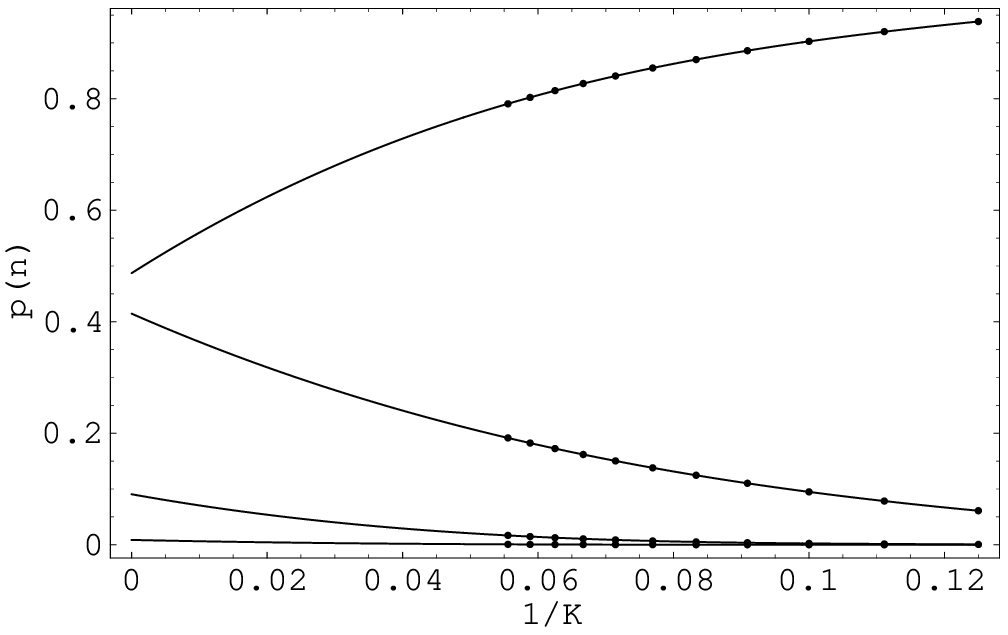,width=8cm}}
\caption{Parton number distribution for the first state
appearing in the spectrum ($M_3^2(\infty)=21.7644$ in the bosonic
$P-$, $S+$ sector). Plotted are
the probabilities $p(n)$ for the state to have $n$ partons vs. $1/K$.
The curves (from top) are for $n=3,5,7,$ and $9$, respectively. }
\label{ppdFits}
\end{figure}

The average number of partons in a bound state only tells part of the
story.  It is necessary to look also at the distribution of partons
and ask how it changes with the
resolution.  We know that the states get longer and longer, but how this
transition happens is
not known. The parton distributions for a particular bound state are
shown in Fig.~\ref{ppd}.  In Fig.~\ref{ppdFits}
we plot fits to the probabilities of finding $n$ partons in this state
as a function of the resolution. From the extrapolation,
we find that the state consists of 48.7\% three-parton, 41.4\%
five-parton, 9\% seven-parton, and 0.9\% nine-parton states. Note that
these results of four independent extrapolations add up almost exactly
to unity.

\section{Counting states}

\subsection{Prime resolutions}
\label{sec:prime}

It is very advantageous to know the exact number of states of a theory in a
discrete approach like SDLCQ, because it may tell us something about 
the physics involved, e.g.~massless eigenstates.
We start by considering the case where the resolution is a prime number,
which is the easiest situation.  Because of the cyclicity of the
color trace that defines the basis states,
there is a many-to-one mapping from sequences of operators to states.
First, we have to find the number of sequences of creation operators
given $n$ species of particles at harmonic resolution $K$.
To make the notation clear, consider the ${\cal N=}(1,1)$ case,
where $n=2$.  A typical sequence of operators at $K=13$ is
\begin{equation}
  s^\dagger = a^\dagger(3)b^\dagger(1)b^\dagger(1)b^\dagger(3)
a^\dagger(1)a^\dagger(4) ,
\end{equation}
where $a^\dagger(k)$ creates a boson with longitudinal momentum
fraction $k/K$ and $b^\dagger(k)$ creates a fermion.
We encode this with the following 13 symbols
\begin{equation}
  s^\dagger = \ a + + \ b \ b \ b + + \ a \ a + +\; + ,
\end{equation}
with the ``$+$`` symbol used to indicate an increase in the
momentum above one; the notation suppresses the dagger symbols. 
The mapping between the two
encodings is one-one and onto, so that their numbers must be the same. 
For a particular value of $K$, the encoding involves picking $K$
sequential choices from a set of three elements, with the restriction
that the first choice cannot be the symbol ``$+$''.  Thus the number of
sequences of operators for ${\cal N}=(1,1)$ is $2\times 3^{K-1}$.  For a
system with $n$ types of operators it is $n(n+1)^{K-1}$.

The number of states is less than the number of operator sequences. 
In our example, traces of the sequences of operators
\begin{eqnarray}
  s_1^\dagger &=& b^\dagger(1)b^\dagger(1)b^\dagger(3)
a^\dagger(1)a^\dagger(4)a^\dagger(3) , \\
  s_2^\dagger &=& b^\dagger(1)b^\dagger(3)a^\dagger(1)
a^\dagger(4)a^\dagger(3)b^\dagger(1) , \nonumber \\
  s_3^\dagger &=& b^\dagger(3)a^\dagger(1)a^\dagger(4)
a^\dagger(3)b^\dagger(1)b^\dagger(1) , \nonumber
\end{eqnarray}
would create the same state as $\Tr[s^\dagger]$, because of the cyclicity
of the trace.  Thus, for the generic
case, a sequence of $j$ operators will have $j-1$ redundant partners
in the sequence listing.  In other words, each sequence of $j$ partons
should only contribute $1/j$ to the number of states.  In
symmetric cases, some of the rotations may not be distinct from the
original sequence.  In that case, the contribution should be greater
than $1/j$.  For $K$ prime, this only occurs for the sequence
$a^\dagger(1)^K$.  We will adjust the final result to account for this. 

To calculate the number of distinct states, we define the following
quantities:
\begin{itemize}
\item $n$ = number of types of operators (not counting ``+'').
\item $K$ = total number of momentum quanta.
\item $j$ = number of partons (symbols other than $+$ in a sequence).
\item $B(K,n,j)$ = number of sequences with given values for $K$, $n$,
and
  $j$.
\item $A(K,n,j)  = \frac{B(K,n,j)}{j}=$ number of $j$-parton states,
  neglecting over-counting due to symmetry.
\item $A(K,n) = \sum_{j=2}^{K}A(K,n,j)$ = number of states.
\end{itemize}
So, if $j = 1$, then $B(K,n,j) = n$. The first symbol must be one of
our $n$ types.  The other $K-1$ symbols are all $+$.  Note that this
case will not be included in the list of states because we explicitly
disallow operators where the momentum is $K$.
In general, we must place one symbol from the $n$ types in the first
position and $j-1$ in the other $K-1$ positions.  These positions
can be selected in 
$\left({\scriptsize\begin{array}{c}  K -1 \\ j -1 \end{array}}\right)$
ways and the $j$ symbol choices selected in $n^j$ ways.  
Therefore, we have
\begin{equation}
  B(K,n,j) =
  n^j
  \left(
    \begin{array}{c}
      K -1 \\ j -1
    \end{array}
  \right). \label{eq16}
\end{equation}
To get the contribution to the number of states, we divide by $j$
\begin{equation}
  A(K,n,j) =
  \frac{n^j}{j}
  \left(
    \begin{array}{c}
      K -1 \\ j -1
    \end{array}
  \right)
  =
  \frac{n^j}{K}
  \left(
    \begin{array}{c}
      K \\ j
    \end{array}
  \right),
\end{equation}
and to obtain the total, we sum over all contributions from the
different parton numbers
\begin{equation}
  A(K,n) =
  \sum_{j=2}^K
  \frac{n^j}{K}
  \left(
    \begin{array}{c}
      K \\ j
    \end{array}
  \right)
  = \frac{1}{K} \left[(1+n)^K - 1 -
  nK
  \right].
\end{equation}
Thus, neglecting symmetry, we have a
formula for the number of states as a function of the number of types
of operators and the harmonic resolution $K$.

For $K$ prime, symmetry only plays a role when $j = K$.  Rotations
of a $K$-parton state, which has only one type of operator, leave the
state unchanged.  Each of the $n$, homogeneous, $K$-parton states
contribute $1/K$ to the above formula and not one.
Our corrected formula for the number of states is therefore
\begin{equation}
  A(K,n)_{\rm prime} =
 \frac{1}{K} \left[(1+n)^K - (1+n) \right].
\end{equation}

\subsection{General resolutions}
\label{sec:generalk}

In this section we extend the above result to
general $K$. We start by defining the functions
\begin{eqnarray}
  \label{eq:C}
C(K,n,1) = n = 2 C_f(K,n,1) = 2 C_b(K,n,1), \\
C(K,n,j) = B(K,n,j) - \sum^K_{q > 1,\, {\rm odd}}
C(\frac{K}{q}, n, \frac{j}{q}),  \nonumber \\
C_f(K,n,j) = B_f(K,n,j) - \sum^K_{q > 1,\, {\rm odd}}
C_f(\frac{K}{q}, n, \frac{j}{q}), \label{17star} \nonumber \\
C_b(K,n,j) = B_b(K,n,j) - \sum^K_{q > 1,\, {\rm odd}}
C_b(\frac{K}{q}, n, \frac{j}{q}). \nonumber
\end{eqnarray}
All functions $C(\frac{K}{q}, n, \frac{j}{q})$ are taken to be zero
when $K/q$ or $j/q$ is not an integer.
The $f$ and $b$ subscripts indicate fermion or boson states,
respectively.
For example, $B_f(K,n,j)$ is the number of sequences of $K$ symbols
with an odd number of fermions and $j$ partons.  We then have
$B_f(K,n,j) = B_b(K,n,j) = \frac{1}{2} B(K,n,j)$, where $B(K,n,j)$ is
given in Eq.~(\ref{eq16}). 
Thus the $C$ functions are related by
$C_f(K,n,j) = C_b(K,n,j) = \frac{1}{2} C(K,n,j)$.

The total number of fermionic states will be the sum over all odd $q
\ge 1$ of fermionic states made from concatenations of sequences of
length $K/q$
\begin{equation}
  A_f(K,n,j) =
  \sum_{q,\,{\rm odd}}  \frac{q}{j}
  C_f\left(\frac{K}{q}, n, \frac{j}{q}\right).
\end{equation}
The factor $q/j$ accounts for the $j/q$ sequences that
map to the same state.  The sum is restricted to odd concatenations
because even concatenations would be bosons and are, in fact, rejected
because of fermion statistics.
Bosonic states can be made from concatenations of any number of
bosonic subsequences.

We define the function $D(K,n,j)$
as the number of symbol sequences of length $K$ that are not
concatenations of smaller sequences.  We have the relation that a
sequence counted in $C(K,n,j)$ may be non-symmetric or it may be a
concatenation of $2^l$ smaller sequences. If it were a concatenation
of $2^l q$ smaller sequences, where $q$ is odd, it would have been
included in $C(K/q,n,j/q)$. Thus we have
\begin{equation} \label{eq:D}
  D(K,n,j) = C(K,n,j) - C\left(\frac{K}{2},n,\frac{j}{2}\right).
\end{equation}
Since $D_f(K,n,j) = C_f(K,n,j)$, we find
\begin{equation}
  D_b(K,n,j) = C_b(K,n,j) - C\left(\frac{K}{2},n,\frac{j}{2}\right)
             = C_b(K,n,j) - 2C_b\left(\frac{K}{2},n,\frac{j}{2}\right).
\end{equation}
Now we are ready to calculate the number of bosonic states
\begin{eqnarray}
  A_b(K,n,j)
  &=&
  \sum_{q}  \frac{q}{j}
  D_b\left(\frac{K}{q}, n, \frac{j}{q}\right) \\
  &=&\sum_{q}
  \frac{q}{j}\left[
  C_b\left(\frac{K}{q}, n, \frac{j}{q}\right) -
  2C_b\left(\frac{K}{2q}, n, \frac{j}{2q}\right)\right]
  =
  \sum^K_{q,\,{\rm odd}}  \frac{q}{j}
  C_b\left(\frac{K}{q}, n, \frac{j}{q}\right). \nonumber
\end{eqnarray}
This result proves that the number of fermions and bosons is equal at
each resolution $K$ and parton number $j$.
We can sum the total number of states over all $j>1$
\begin{equation} \label{eq:AfKn}
  A_f(K,n) =
  \sum_j
  \sum_{q,\,{\rm odd}}  \frac{q}{j}
  C_f\left(\frac{K}{q}, n, \frac{j}{q}\right)
  -  A_f(K,n,1)
  =
  \sum_{q, \,{\rm odd}}
  A^{\prime}_f\left(\frac{K}{q},n\right) - \frac{n}{2},
\end{equation}
where we have defined the non-symmetric state number function
\begin{eqnarray}
  A^{\prime}_f(K,n) &\equiv& \sum_j \frac{1}{j} C_f(K,n,j) =
  \sum_j \frac{1}{j} B_f(K,n,j) -\sum_{q>1, \,{\rm odd}}
\sum_j\frac{1}{j}
  C_f\left(\frac{K}{q}, n,\frac{j}{q}\right)\nonumber\\
 &=& \sum_j \frac{1}{j}B_f(K,n,j) -\sum_{q>1, \,{\rm odd}}
  \frac{1}{q} A^{\prime}_f\left(\frac{K}{q}, n\right)\nonumber\\
  &=&\frac{(1+n)^K-1}{2K}-\sum_{q>1,\, {\rm odd}}
  \frac{1}{q} A^{\prime}_f\left(\frac{K}{q}, n\right).
\label{Afprime}
\end{eqnarray}
The result for $A_b(K,n)$ is evidently the same, and we
have $A(K,n) = 2A_f(K,n)$.

\subsection{Symmetry sectors}
\label{sec:symmetry}

There are several symmetries of
the Hamiltonian which can be used to reduce the numerical effort by 
block diagonalization. Each block corresponds to a distinct symmetry
sector of the theory.
In the ${\cal N}=(1,1)$ theory these are supersymmetry,
$S$ symmetry (\ref{Z2-S}), and parity (\ref{Z2-P}). 
Thus the ${\cal N}=(1,1)$ theory has eight symmetry sectors.
The supercharge $Q^-$ changes the fermion number by one,
taking a sector to its image sector. 
In many cases the image sector will have fewer states,
which requires the existence of massless states in the initial sector,
since massless states are annihilated by $Q^-$.
Knowledge of the number of states in
each sector can provide a lower bound on the number of massless states.

The ${\cal N}=(2,2)$ theory has an additional $Y$ symmetry:
\begin{equation}\label{Z2-Y}
Y : a_1(k^+)\leftrightarrow a_2(k^+), \qquad
    b_1(k^+)\rightarrow b_1(k^+), \qquad
      b_2(k^+)\rightarrow -b_2(k^+).
\end{equation}
The eight sectors determined by
supersymmetry, parity, and $S$-symmetry are each split exactly in half
by this additional symmetry.  Thus the minimum number of massless states 
is not changed if $Y$ symmetry is present.
 
In this section, we will provide formulas
for the numbers of states in the four supersymmetry and parity sectors.  
We will show that the total number of positive and
negative eigenvalues for the $Y$ symmetry are equal.  
Then we shall show that the $SY$ operation has the same number of positive 
and negative eigenvalues in each of the four supersymmetry and parity 
sectors and calculate the number of additional positive eigenvectors 
that the $S$ symmetry possesses.  We then apply the results to find the 
number of states for each sector.  Finally, we apply the count in each sector
to the problem of counting massless states.

\subsubsection{Supersymmetry and parity}

{}From the previous section we know that
\begin{equation}
  A_{fo}(K,n) = A_{bo}(K,n) \qquad \mbox{and} \qquad
  A_{fe}(K,n) = A_{be}(K,n),
\end{equation}
where the subscripts denote fermion and boson sectors
and odd and even parton numbers, respectively.
$A_{fo}$ and $A_{be}$ count states with even parity and are taken into
each other by 
$Q^-$ which changes the fermion number of a state by one
and changes the boson number by two or none.
In order to compare numbers of states, we need therefore to
relate the counts of even and odd-parton states.

We consider the number of fermionic sequences
$B_f(K,n,j)=\frac{1}{2}B(K,n,j)$;
see Eq.~(\ref{eq16}).
In this function, the power of $n$ is even if and only if $j$ is even. 
Next we consider the number of non-symmetric fermionic sequences,
given in Eq.~(\ref{17star}).
Since $q$ is restricted to odd values, $j/q$ will be even (odd)
when $j$ is even (odd) and therefore, by induction, $B_f(K,n,j)$
will have even (odd) powers of $n$ when $j$ is even (odd).
In fact, the even-odd dependence carries through the whole calculation.
Therefore, we have
\begin{eqnarray}
  \label{eq:Abe}
  A_{be}(K,n) = A_{fe}(K,n) = \frac{1}{2}\left[A_f(K,n) +
A_f(K,-n)\right],
\\
  A_{bo}(K,n) = A_{fo}(K,n) = \frac{1}{2}\left[A_f(K,n) -
A_f(K,-n)\right].
\nonumber
\end{eqnarray}
Thus we have separated the states into four sectors, and we know the
numbers of states in each sector.

\subsubsection{$Y$ symmetry}

For ${\cal N}=(2,2)$ SYM theory, the situation is
different, because our basis states are not all eigenstates of the
additional $Y$ symmetry.  Most Fock states $\Ket{\psi}$ are mapped onto
distinct states $Y\Ket{\psi}$ by the symmetry.  A $Y$-even (odd)
eigenvector will then be of the form $\Ket{\psi} \pm Y\Ket{\psi}$.
Thus, if each state had distinct partner under $Y$, the space would clearly
split exactly in half between states with positive and negative eigenvalues.
However, there are states such that $Y\Ket{\psi} = \pm \Ket{\psi}$, and
they must be counted before showing that the space does indeed
split exactly in half.

There are two ways that a Fock state can be an eigenstate of $Y$.
The first is for it to have no bosons.  In that case, the $Y$ 
eigenvalue is determined by the number of $b_2^\dagger$ operators.  
The second is for the state to have the
form $\Tr\left[(s^\dagger\bar s^\dagger)^q\right]\Ket{0}$, where
$s^\dagger$ is a sequence of creation operators, $\bar s^\dagger$ is
the sequence obtained from $s^\dagger$ by interchanging $a_1^\dagger$
and $a_2^\dagger$, and $q$ is a positive integer.   The
sequences of this type will have an even number of $b_2^\dagger$
operators.  Their $Y$ eigenvalue is determined by the fermion number 
of $s^\dagger$, unless $s^\dagger=\bar s^\dagger$, which will occur
for purely fermionic sequences.

The problem of counting the positive and negative $Y$ eigenvalues
among the states that have no bosons in the case $n=4$ is almost
equivalent to the counting of fermion and boson states in the case 
$n=2$.  We map the $n=4$ sequences to $n=2$ according to
the replacements $b_1^\dagger\rightarrow a^\dagger$ and
$b_2^\dagger\rightarrow b^\dagger$.  However, due to fermion
statistics, there is not a perfect one-to-one correspondence
between $n=2$ states and purely fermionic $n=4$ states.
In particular, $n=4$ states created from an even number of
even parton number subsequences, which each have an odd number of
$b_2^\dagger$ operators, will have no corresponding $n=2$ state.
For resolution, $K$, and $j$ partons, the count of these states, 
which all have positive $Y$ eigenvalue, is
\begin{equation}
\label{eq:Y1plus}
\delta Y_{1+}(K,j)=\sum_{q,\,{\rm even}} 
      \frac{q}{j} C_f\left(\frac{K}{q},2,\frac{j}{q}\right).
\end{equation}
Also, some $n=2$ boson states have no corresponding $n=4$ states because
they would consist of an even number of odd parton number subsequences.
The corresponding $n=2$ states of this latter type have as their total
\begin{equation}
\label{eq:Y1minus}
\delta Y_{1-}(K,j)=
\sum_{\scriptsize \begin{array}{c} j/q,\,{\rm odd} \\ 
                       q,\,{\rm even} \end{array}}
\frac{q}{j} D\left(\frac{K}{q},2,\frac{j}{q}\right)
=
\sum_{\scriptsize \begin{array}{c} j/q,\,{\rm odd} \\ 
                  q,\,{\rm even} \end{array}}
\frac{q}{j} \left[C\left(\frac{K}{q},2,\frac{j}{q}\right)
     -C\left(\frac{K}{2q},2,\frac{j}{2q}\right)\right].
\end{equation}
The last term, $C\left(\frac{K}{2q},2,\frac{j}{2q}\right)$,
is zero, because $j/q$ is odd and therefore not divisible by 2. 

The difference between (\ref{eq:Y1plus}) and (\ref{eq:Y1minus}) 
gives the difference between $n=4$ purely fermionic positive-$Y$ 
states and $n=2$ bosonic states.  Since $n=2$ has an equal number of
boson and fermion states, the net difference between numbers of
positive and negative $Y$ eigenvalues for purely fermionic states 
with $j$ partons is
\begin{eqnarray}
\delta Y_{1+}(K,j)-\delta Y_{1-}(K,j)
&=&\sum_{q,\,{\rm even}} 
     \frac{q}{j}C_f\left(\frac{K}{q},2,\frac{j}{q}\right)
  -\sum_{\scriptsize \begin{array}{c} j/q,\,{\rm odd} \\ 
                 q,\,{\rm even} \end{array}}
    \frac{q}{j} 2C_f\left(\frac{K}{q},2,\frac{j}{q}\right) \nonumber \\
&=& \sum_{q,\,{\rm even}} 
    (-1)^{j/q}\frac{q}{j}C_f\left(\frac{K}{q},2,\frac{j}{q}\right).
\end{eqnarray}

We next consider the contribution from $Y$ eigenstates of the form
$\Tr\left[(s^\dagger\bar s^\dagger)^q\right]\Ket{0}$.   To avoid
duplication, we count only those sequences $s^\dagger\bar s^\dagger$ 
that cannot be further subdivided as 
$(s^{\prime\dagger} \bar s^{\prime\dagger})^{(2r+1)}$ for some
sequence $s^{\prime\dagger}$ and some integer $r>0$.  For $j$
partons and resolution $K$, the number of such states is
$\frac12 \frac{2q}{j} C\left(\frac{K}{2q},n,\frac{j}{2q}\right)$,
where we have included a factor of 1/2 to account for the fact
that $\Tr\left[(s^\dagger\bar s^\dagger)^q\right]\Ket{0}$ is not
distinct from $\Tr\left[(\bar s^\dagger s^\dagger)^q\right]\Ket{0}$.
Since $C_b=C_f$, there are equal numbers of positive and
negative eigenvalues among these states.  However, the purely
fermionic states have already been counted and must be removed from
this second contribution.  They are subtracted from the positive count 
when $j/2q$ is even and from the negative count when it is odd.
Each count is done as before, by mapping to $n=2$ states,
and the net contribution is
\begin{equation}
\delta Y_{2+}(K,j)-\delta Y_{2-}(K,j)
  =-\sum_{q\geq 1} (-1)^{j/2q}\frac{q}{j}
                C\left(\frac{K}{2q},2,\frac{j}{2q}\right).
\end{equation}
Since $C=2C_f$, this exactly cancels 
$\delta Y_{1+}(K,j)-\delta Y_{1-}(K,j)$.

\subsubsection{$S$ and $SY$ symmetry} \label{sec:SSY}

The $S$ symmetry swaps the color indices of the operators and
multiplies each by $-1$.  It
is equivalent to reversing the order of the operators in a state
and multiplying by a sign $(-1)^j(-1)^{\frac{n_f(n_f-1)}{2}}$,
where $j$ is the number of partons and
$n_f$ is the number of fermions in the state. 
In this section we look for Fock states that are eigenstates of $S$
or, for the ${\cal N}=(2,2)$ theory, of $SY$. 
We look for the difference between
the numbers of positive and negative eigenvalues of these two operators.
Most of these eigenvectors come in pairs, one with positive and one
with negative eigenvalue but with equal parton number and the same
parity of fermion number.  We only need to count the ones that do not
come in pairs.
For Fock states $\Ket{\psi}$ which are not eigenstates we again
form combinations $\Ket{\psi}\pm S\Ket{\psi}$ or $\Ket{\psi}\pm
SY\Ket{\psi}$
as pairs of eigenstates with opposite eigenvalues.  For Fock states
which {\em are} eigenstates, we find only pairs with opposite
eigenvalues except for two-parton states.

We can see this by considering even and odd parton number
separately.  For odd parton number, we rotate the sequence of symbols
to the form $s^\dagger x^\dagger \bar s_R^\dagger$ where the sequence
$s_R^\dagger$ is the reversal of $s^\dagger$ and
$\bar s_R^\dagger=s_R^\dagger$ or $Y s_R^\dagger$,
depending on whether we are considering the $S$ symmetry or
the $SY$ symmetry, and $x^\dagger$ is a
single operator satisfying $Yx^\dagger = \pm x^\dagger$ under $SY$ 
symmetry. By changing the first operator in $s^\dagger$ from boson to 
fermion, we change the fermion number by two and the sign of $S$ while 
changing neither parton number nor fermion-number parity.
Thus states of this form do come in pairs.

For even parton number the argument is more complicated, and we just
give a sketch.  There is a two-to-one map between sequences of operators
of the types $s^\dagger \bar s_R^\dagger$ and
$x^\dagger s^\dagger y^\dagger \bar s_R^\dagger$ and basis states that
are eigenvalues of $S$ or $SY$, where $s^\dagger$ and $\bar s_R^\dagger$ 
are subsequences and $x^\dagger$ and $y^\dagger$ are single operators, 
as before.  Again, we can flip the first operator in $s^\dagger$ from 
fermion to boson to swap the sign of the $S$ or $SY$ eigenvalue.  Some
states are absent due to fermion statistics, but they also come in pairs.

The argument does not apply to two-parton states, which we now
handle explicitly.  All two-parton basis states are eigenstates of $S$
with positive eigenvalue.  All $S$ does in this case is swap the dummy
summation indices.  The number of two-parton fermionic states is
$(\frac{n}{2})^2(K-1)$.  This comes from two choices of which fermion
and boson to include and $K-1$ choices as to how to apportion the
momentum between the two operators.  The number for two-parton bosonic
states is the same.

The two-parton Fock states that are eigenstates of $SY$ are listed
in Table~\ref{tab:SY}; positive and negative eigenvalues appear
there in equal numbers.  The two-parton fermion states have 
exactly one boson operator and cannot be eigenstates of $Y$. 
The numbers of positive and negative eigenvalues of $SY$ are
therefore equal.
\begin{table}[t]
\centerline{
\begin{tabular}{cccc}
\hline \hline
    &  &  \multicolumn{2}{c}{number of states} \\
  State & $SY$ & $K$ even & $K$ odd \\
\hline
$\Tr\left[a_1^\dagger(K/2) a_2^\dagger(K/2)\right]\Ket{0}$ & 1 & 1 
                                        & 0 \\
$\Tr\left[b_1^\dagger(k) b_2^\dagger(K-k)\right]\Ket{0}$ & $-1$ & $K-1$ 
                                        & $K-1$ \\
$\Tr\left[b_1^\dagger(k) b_1^\dagger(K-k)\right]\Ket{0}$ & 1 & $K/2-1$ 
                                        & $(K-1)/2$ \\
$\Tr\left[b_2^\dagger(k) b_2^\dagger(K-k)\right]\Ket{0}$ & 1 & $K/2-1$ 
                                        & $(K-1)/2$ \\
\hline \hline
\end{tabular}}
\caption{Two-parton Fock states that are eigenstates of $SY$.
\label{tab:SY}}
\end{table}
The number of $S+$ and $S-$ eigenstates is equal for odd parity and even
or odd fermion numbers, but in the even-parity fermion and
boson sectors, the number of $S+$ eigenstates is greater than the
number of $S-$ eigenstates by $(\frac{n}{2})^2(K-1)$.

\subsubsection{Results}

We label the $Y$ and $S$ symmetry
sectors by additional letters  $y^\pm$ and $s^\pm$, e.g.
$A_{bey^+} + A_{bey^-} = A_{be}$. It is obvious that
no odd-parity Fock state, with its odd number of bosons,
can be an eigenstate of $Y$ or $SY$; therefore,
$Y$ and $SY$ eigenvalues occur in equal numbers in each
such sector, and we have
\begin{eqnarray}
  A_{fes^+y^+} =
  A_{fes^+y^-} \qquad \mbox{and}\qquad
  A_{fes^-y^+} =
  A_{fes^-y^-}, \\
  A_{bos^+y^+} =
  A_{bos^+y^-} \qquad \mbox{and}\qquad
  A_{bos^-y^+} =
  A_{bos^-y^-}.   \nonumber 
\end{eqnarray}
Fermionic states with an odd number of partons can be eigenstates, but
they come in pairs related by the exchange of $b_1$ and $b_2$.  Therefore, the
counts in these sectors must also be equal:
\begin{eqnarray}
  A_{fos^+y^+} =
  A_{fos^+y^-} \qquad \mbox{and}\qquad
  A_{fos^-y^+} =
  A_{fos^-y^-}.
\label{star}
\end{eqnarray}
The bosonic even-parity sectors are all that remain to be considered. 
Since we have $A_{y^+} = A_{y^-}$, the totals for the bosonic even-parton
sector must also be equal:
\begin{eqnarray}
  A_{bey^+} = A_{bey^-} .
\end{eqnarray}
Furthermore, since $SY$ has an equal number of positive and negative
eigenvalues, we have 
\begin{eqnarray}
  A_{bes^+y^+} + A_{bes^-y^-}  = 
  A_{bes^+y^-} + A_{bes^-y^+},
\end{eqnarray}
and, therefore,
\begin{eqnarray}
  A_{bes^+y^+} =
  A_{bes^+y^-} \qquad \mbox{and}\qquad
  A_{bes^-y^+} =
  A_{bes^-y^-}.
\end{eqnarray}

For eigenstates of $S$ alone, we have
\begin{eqnarray}
  A_{bes^+} -   A_{bes^-} =
  A_{fes^+} -   A_{fes^-}&=&\frac{n^2}{4}(K-1)\\
  A_{bos^+} -   A_{bos^-} =
  A_{fos^+} -   A_{fos^-}&=&0 . \nonumber
\end{eqnarray}
Thus, from Eq.~(\ref{eq:Abe}) we obtain
\begin{eqnarray}
  A_{bes^+}(K,n) = A_{fes^+}(K,n) &=&
\frac{1}{4}\left\{A_f(K,n) + A_f(K,-n) + \frac{n^2}{2}(K-1)\right\}, \\
  A_{bes^-}(K,n) = A_{fes^-}(K,n) &=& \frac{1}{4}
\left\{A_f(K,n) + A_f(K,-n) -\frac{n^2}{2}(K-1) \right\},  \nonumber \\
  A_{bos^+}(K,n) = A_{fos^+}(K,n) &=&
  \frac{1}{4}\left\{A_f(K,n) - A_f(K,-n)\right\} \nonumber \\
&=&A_{bos^-}(K,n) = A_{fos^-}(K,n). \nonumber
\end{eqnarray}

We present the results for the numbers of states in the different
sectors for the ${\cal N}=(1,1)$ and
${\cal N}=(2,2)$ theories in Tables~\ref{Numbers1} and \ref{Numbers2},
respectively. 
\begin{table}[t]
\begin{center}
\begin{tabular}{cccc}
\hline \hline
$K$ &$ A_{bes^+} $&$ A_{bos^+} $&$ A_{bes^-} $\\
\hline
2  & 1 & 0 & 0 \\
3  & 2 & 1 & 0 \\
4  & 4 & 2 & 1 \\
5  & 8 & 6 & 4 \\
6  & 18 & 15 & 13 \\
7  & 42 & 39 & 36 \\
8  & 106 & 102 & 99 \\
9  & 278 & 274 & 270 \\
10  & 743 & 738 & 734 \\
11  & 2018 & 2013 & 2008 \\
12  & 5543 & 5537 & 5532 \\
13  & 15336 & 15330 & 15324 \\
14  & 42712 & 42705 & 42699 \\
15  & 119586 & 119579 & 119572 \\
16  & 336310 & 336302 & 336295 \\
17  & 949568 & 949560 & 949552 \\
18  & 2690439 & 2690430 & 2690422 \\
19  & 7646466 & 7646457 & 7646448 \\
20  & 21792414 & 21792404 & 21792395 \\
\hline \hline
\end{tabular}
\end{center}
\caption{Number of states in the different symmetry sectors
for the ${\cal N}=(1,1)$ theory. Recall that $A_{bes^\pm}= A_{fes^\pm}$
and $A_{bos^+}= A_{fos^+}=A_{bos^-}= A_{fos^-}$.\label{Numbers1}}
\end{table}
\begin{table}[t]
\begin{center}
\begin{tabular}{cccc}
\hline \hline
$K$ &$ A_{bes^+} $&$ A_{bos^+} $&$ A_{bes^-} $\\
\hline
2  & 4 & 0 & 0 \\
3  & 8 & 6 & 0 \\
4  & 28 & 16 & 16 \\
5  & 80 & 84 & 64 \\
6  & 352 & 310 & 332 \\
7  & 1368 & 1434 & 1344 \\
8  & 6220 & 6000 & 6192 \\
9  & 26872 & 27404 & 26840 \\
10  & 122828 & 121332 & 122792 \\
11  & 552872 & 556878 & 552832 \\
12  & 2548704 & 2537606 & 2548660 \\
13  & 11722224 & 11752860 & 11722176 \\
14  & 54538408 & 54452970 & 54538356 \\
15  & 254193656 & 254432786 & 254193600 \\
16  & 1192429228 & 1191756592 & 1192429168 \\
17  & 5608899392 & 5610798480 & 5608899328 \\
18  & 26493643916 & 26488263020 & 26493643848 \\
19  & 125475816264 & 125491109142 & 125475816192 \\
20  & 596068240212 & 596024655364 & 596068240136 \\
\hline \hline
\end{tabular}
\end{center}
  \caption{Number of states in the different symmetry sectors
for the ${\cal N}=(2,2)$ theory. Recall that $A_{bes^\pm}= A_{fes^\pm}$
and $A_{bos^+}= A_{fos^+}=A_{bos^-}= A_{fos^-}$.\label{Numbers2}}
\end{table}
How can we make use of these results? Since the
supercharge $Q^-$ maps one symmetry sector to another, we can conclude
that if one sector has $r$ more states than the other, then the Hamiltonian
$P^-\propto (Q^-)^2$ has at least $r$ massless states in that sector.

In the ${\cal N}=(1,1)$ theory, we can calculate $A_f(K,-2)$ explicitly.
{}From Eq.~(\ref{Afprime}) we know that $A^\prime_f(1,-2)=-1$ and
$A^\prime_f(2,-2)=0$. Then for $A'_f(2K,-2)$ the recursive sum
yields zero.  The net result is the same for $A'_f(2K+1,-2)$:
\begin{equation}
  A^\prime_f(2K+1,-2) = \frac{1}{4K+2}\left\{[1 + (-2)]^{2K+1} - 1 -
  2 A^\prime_f(1,-2)\right\} = 0.
\end{equation}
{}From these and Eq.~(\ref{eq:AfKn}) we have
\begin{eqnarray}
  A_f(2K+1,-2) &=& A^\prime_f(2K+1,-2) + A^\prime_f(1,-2) -  (-1) = 0 ,
\\
  A_f(2K,-2) &=& A^\prime_f(2K,-2)  -  (-1) = 1. \nonumber
\end{eqnarray}
This means that for the ${\cal N}=(1,1)$ case, 
the bosonic and fermionic parity-even and $S$-even sectors must have
$(K-1)/2$ (for odd $K$) or $K/2$ (for even $K$) more states than their
parity-odd counterparts. Likewise,
the bosonic and fermionic parity-odd and $S$-odd sectors must have
$(K-1)/2$ (odd $K$) or $K/2-1$ (even $K$)
more states than their parity-even counterparts.
The minimum number of massless states is therefore $2(K-1)$, regardless
of whether the resolution $K$ is even or odd.

\section{Discussion}
\label{sec:discussion}
%
In this paper we discussed a number of properties
of ${\cal N}=(1,1)$ super Yang--Mills theory in two dimensions at large
$N_c$ in a SDLCQ approach. We
found the low-energy bound-state spectrum in each of the symmetry sectors
and extrapolated the spectrum to obtain the continuum masses.
Although convergence is fast and approximately linear, we were careful
to push our numerical calculations to large resolution $K$.
We found that this spectrum exhibits an interesting distribution
of masses. We explained this pattern and
discussed how one can understand it by constructing a toy
model, that one might use to approximate the actual spectrum.
We looked at some of the properties of a number of these bound states
in detail.

Finally, we considered the numbers of states in the Fock basis in a SDLCQ
approximation. Obviously, one would like to know how many basis
states one will encounter before one attempts a numerical approximation, 
since, in an actual SDLCQ calculation, we
solve the problem separately in each symmetry sector. We calculated 
the number of states in each symmetry sector algebraically.
It is useful to know which sector has the smallest number of states,
because this sector will generally have the
smallest number of massless states, which makes it the simplest one to
solve numerically. Our calculation showed that the number of
fermion states is exactly equal to the number of boson states. This
represents part of the answer as to why the SDLCQ
approximation is exactly supersymmetric.

The results presented in this paper nearly complete the solution of
the ${\cal N}=(1,1)$ super Yang--Mills theory 
in two dimensions. In previous
papers~\cite{Antonuccio:1998kz,Hiller:2004vy}
we had calculated the thermodynamics of the model and the behavior of
the correlator of the energy momentum tensors and the distribution
function of some of the bound states.  We have
discussed the spectrum of the theory and the counting of its
states in this paper.
In terms of the general behavior of this theory, little more than
the analysis of the massless states remains to be done.

\section*{Acknowledgments}
This work was supported in part by the U.S. Department of Energy
and by the Minnesota Supercomputing Institute. One of the authors (SP)
would like to
acknowledge the Aspen Center for Physics where this work was completed.
The research of Uwe Trittmann was supported by an award from the
Research Corporation.

\end{document}